\documentclass{article}%
\input epsf
\usepackage{amsmath}%
\usepackage{amsfonts}%
\usepackage{amssymb}%
\usepackage{graphicx}
\newcommand{\be}[1]{
\begin{eqnarray}\label{#1}}
\newcommand{\ee}{\end{eqnarray}}
\newcommand{\ci}[1]{\cite{#1}}
\newcommand{\re}[1]{(\ref{#1})}

\newcommand{\FF}{F^2_\pi}
\newcommand{\ba}{\begin{array}}
\newcommand{\ea}{\end{array}}

\newcommand{\partialboth}{\stackrel{\leftrightarrow}{\partial}}
\newcommand{\chil}[1]{\stackrel{o}{#1}}

\newcommand{\insertfig}[2]{\mbox{\epsfxsize=#1cm \epsfbox{#2.eps}}}

\begin{document}


\begin{center}
{\Large Large-N Summation of  Chiral Logs for Generalized Parton Distributions}\\[0.5cm]

 N. Kivel$^{a,b}$, M.V.~Polyakov$^{a,b}$, A.~Vladimirov$^{b,c}$\\[0.3cm]

\footnotesize\it $^a$ Petersburg
Nuclear Physics Institute, Gatchina, St. Petersburg 188350,
Russia\\
 \footnotesize\it $^b$ Institute for
Theoretical Physics II, Ruhr University Bochum, 44780 Bochum, Germany
\\
\footnotesize\it $^c$
Bogolubov Laboratory of Theoretical Physics, JINR, 141980 Dubna, Russia

\vspace*{0.44cm}

\end{center}

\begin{abstract}
We demonstrate that in the region of Bjorken $x_{\rm Bj}\sim
m_\pi^2/(4\pi F_\pi)^2$ and/or $x_{\rm Bj}\sim |t|/(4\pi F_\pi)^2$
the standard $\chi$PT for the pion  GPDs fails and one
must perform all order resummation of $\chi$PT. We
perform such resummation in the large-$N$ limit of the $O(N+1)$ extension of the chiral theory. Explicit resummation
allows us to reveal novel phenomena
-- the form of the leading chiral correction to pion PDFs and GPDs
depends on the small $x$ asymptotic of the pion PDFs. In particular, if the pion PDF in the chiral limit has the
Regge-like small $x$ behaviour $q(x)\sim 1/x^\omega$,
the leading large impact parameter
($b_\perp\to\infty$) asymptotics of the quark distribution in the
transverse plane has the form ($m_\pi=0$)
$q(x,b_\perp)\sim 1/x^\omega\ \ln^{\omega}(b_\perp^2)/b_\perp^{2{(1+\omega)}}$.
This result is model independent and
it is controlled completely by the all order resummed $\chi$PT developed in this paper. This asymptotic
interweaves with small-$x$ behaviour of usual PDFs, hence it  depends on the scale,
 at which the corresponding PDF is defined. This is a new and interesting
result in which the chiral expansion meets the QCD evolution.

\end{abstract}

\section*{Introduction}

The generalized parton distributions (GPDs) \cite{pioneers}
(see Refs.~\cite{GPV,Diehlrev,Belitskyrev,Boffi} for recent reviews)
are ditermined by the matrix elements between hadronic states of the well-defined
QCD quark-gluon operators of the type~:
\begin{equation}
 O(\lambda)  =
 \bar{q}\left( {\textstyle \frac{1}{2}}\lambda~n\right)
\gamma_{+}\
q\left(  -{\textstyle \frac{1}{2}}\lambda~n\right)  ,
\label{Oper}
\end{equation}
which is defined on the light-cone, i.e. $n^2=0$ (the gauge link between different points is assumed).
The dependence of GPDs\footnote{Usual parton distributions are just forward limit of GPDs.}
on soft momenta and/or pion mass can be controlled by Chiral Perturbation Theory ($\chi$PT),
see Refs.~\cite{sav,kiv02,che,man,Ando,Dorati}. However, in the case of $\chi$PT for GPDs
apart from external soft momenta scale,
one has to take into account
 another momentum scale in the problem -- the inverse light-cone distance
$1/\lambda$ in the operator (\ref{Oper}). When these two scales are of the same chiral order $\chi$PT requires resummation \cite{my}.

In Ref.~\cite{my} it was shown that the chiral
expansions
of the pion PDFs  $q(x)$ (isovector) and $Q(x)$ (isosinglet)  can be written in leading logs as:
\be{QchptD}
Q(x)  &
=& Q^{reg}(x)+\sum_{n\geq 2,\, {\rm even}}D_n\, \left[a_\chi\ \ln\left(\frac{1}{a_\chi}\right)\right]^{n} \delta^{(n-1)}
(x)\, ,
\\
q(x)  & =& q^{reg}(x)+\sum_{n\geq 1,\, {\rm odd}}D_n\, \left[a_\chi\ \ln\left(\frac{1}{a_\chi}\right)\right]^{n}
\delta^{(n-1)}(x)\, ,
\label{qchptD}%
\ee
where $a_\chi=\left(m_\pi/4\pi F_\pi~\right)^{2}$
 is the chiral
expansion parameter, $F_\pi\approx 93$~MeV is the pion decay constant.
The superscript $reg$ denotes the regular contributions which do not contain the $\delta-$functions, the summation
index $n$ corresponds to the number of loops in $\chi$PT.
\footnote{
In Ref.~\cite{my} the coefficients $D_{1,2,3}$
were computed performing three-loop calculations with the result:
\be{3loops}
D_1=-1, ~~D_2=-\frac 5 3\langle x \rangle,~~ D_3=-\frac{25}{108} \langle x^2 \rangle\, .
\ee}
The presence of the derivatives of the $\delta-$functions requires the reorganization of the chiral expansion
because for $x\sim a_\chi$ all terms in above sums are of the same chiral order.

For the case of GPDs the problem is even more obvious. It was
demonstrated in Ref.~\cite{my} that GPDs contain singular
contributions in the kinematical variable $\xi$ ($\xi=
x_{\rm Bj}/(2-x_{\rm Bj})$). For example, appearance  of such singular terms
in GPDs leads to the following contributions to the leading order
amplitude for hard exclusive processes\footnote{For simplicity we
consider the chiral limit $m_\pi=0$.}: \be{amplitude} {\cal
A}(\xi,t)=\int_{-1}^{1}dx\ \frac{H(x,\xi,t)}{\xi-x}={\cal
F}^{reg}(\xi,t)+\sum_{k=1}^\infty {\cal A}_{k}\ \frac{1}{\xi^{k}}\
\left[{b_\chi\ln(1/b_\chi)}\right]^k\, , \ee where we introduced
small expansion parameter $b_\chi=|t|/4(4\pi F_\pi)^2$. The
presence of such strong singularities $\sim 1/\xi^{k}$ in
region of small $\xi$  compensates
the smallness of chiral expansion parameter and therefore standard
chiral expansion fails. Note that in the case of the nucleon
target such failure of the standard chiral expansion is
imperative, because if $t\sim O(p^2)$ in the chiral counting than
$\xi\sim O(p^2)$ by kinematical constraints.
We see clearly that one has to perform the resummation of
singular contributions in order to obtain correct chiral expansion of GPDs and PDFs.

In present paper we perform such resummation in the large-$N$ limit, where $N$ is the
number of Goldstone bosons ($N=3$ in real world).
The chiral Lagrangian for pions is equivalent to $O(4)$ $\sigma$-model
because of the homeomorphism  $SU(2)\otimes SU(2)=O(4)$.
The extension to arbitrary $N$ is done by the generalization $O(4)\to O(N+1)$.
The $O(N+1)$ $\sigma-$model in the large $N$ limit can be solved by the semiclassical
methods (see Appendix~B for details on $1/N$ expansion).
In such approach one performs partial summation of $\chi$PT diagrams to all orders in accordance with the large-$N$ counting rules.
We stress that that this partial summation respects the chiral symmetry.

\section*{Large$-N$ extension of chiral Lagrangian and resummation of $\chi$PT diagrams }

We  start with the description of the large$-N$ extension of the standard $\chi$PT.
The leading order chiral Lagrangian  describing the low-energy dynamics of pions reads \cite{weinberg}
\be{L2}
\mathcal{L}_{2}=\frac{F_\pi^{2}}{4}\text{~} \text{tr}\left[ \left(  \partial_\mu U\partial_\mu
U^{\dagger}\right)  +m_{\pi}^{2}\left(  U+U^{\dagger}\right)  \right]
\ee
where $F_\pi\approx 93~\text{MeV}$.
We write in the chiral Lagrangian the physical values of the pion mass and decay constant because
their difference with the corresponding bare constants is irrelevant for the leading log approximation we use in the present paper.
The $SU(2)$ matrix $U$ in Eq.~(\ref{L2}) is parameterized in the form:
\be{def:U}
U=\frac{1}{F_\pi}(~\sigma+i~\pi\cdot\tau),~~\ \sigma=F_\pi\sqrt{1-\pi^{2}/F_\pi^{2}}%
\ee
which gives the following result for the chiral Lagrangian:
\be{L:O3}
\mathcal{L}_{2}=\text{~}\frac{1}{2}\left[  \partial_{\mu}\sigma\partial_{\mu
}\sigma+\partial_\mu\pi^{a}\partial_\mu\pi^{a}\right]  +\sigma~F_\pi~m_{\pi}^{2}%
,~\ \ \sigma^{2}+\sum_{a=1}^3\pi^{a}\pi^{a}=F_\pi^{2}%
\ee
We have obtained the Lagrangian of $O(4)$ $\sigma-$model.
Next we extend this model replacing the group  $O(4)$ with $O(N+1)$,
where $N$ has a meaning of the number of Goldstone bosons:
\be{L:ON}
\mathcal{L}_{2}=\text{~}\frac{1}{2}\left[  \partial_{\mu}\sigma\partial_{\mu
}\sigma+\partial_{\mu}\pi^{a}\partial_\mu\pi^{a}\right]  +\sigma~F_\pi~m_{\pi}^{2}%
,~\ \ \sigma^{2}+\sum_{a=1}^{N}\pi^{a}\pi_{a}=F_\pi^{2}%
\ee
Substituting the $\sigma-$field in terms of $\pi$-fields  we obtain%
\be{L:ONpi}
\mathcal{L}_{2} &  =\text{~}-\frac{1}{2}\pi^{a}\partial^{2}\pi^{a}+\frac{1}%
{2}\frac{\left(  \pi^{a}\partial_{\mu}\pi^{a}\right)  ^{2}}{\left(  F_\pi^{2}%
-\pi^{2}\right)  }+\sqrt{F_\pi^{2}-\pi^{2}}~F_\pi m_{\pi}^{2}\\
&  =-\frac{1}{2}\pi^{a}\partial^{2}  \pi
^{a}+N~V(\pi^{2}/N)\label{O_N}%
\ee
where we assume the sum over the repeated indexes of the pion fields and
\be{def:V}
V(\pi^{2}/N)  & =&\frac{1}{8}\frac{\left(  \partial_{\mu}\pi^{2}/N\right)
^{2}}{\left(  G_\pi^{2}-\pi^{2}/N\right)  }+\sqrt{G_\pi^{2}-\pi^{2}/N}~G_\pi~m_{\pi}%
^{2},~\\
G_\pi^{2}  & =&F_\pi^{2}/N~.
\ee
The Lagrangian (\ref{L:ONpi}) is  written in the form which is convenient
for the construction of the large$-N$ perturbation theory. Note that large$-N$ expansion
implies that  the introduced coupling $G_\pi$ is of order $N^0$ with respect to $N$, i.e.
chiral coupling $F_\pi$ scales as $F_\pi\sim \sqrt N$ in the large $N$ limit.

Now, we have to introduce the generalization for the operators $O^{(L,R)}(\lambda)$ (\ref{matchingL},\ref{matchingR}) that
define GPDs and PDFs, see Appendix~A for all definitions and conventions. The isovector operator in $O(4)-$case has the form:
\be{Op:O3}
O^{c}(\lambda)  & =-i\varepsilon\lbrack abc]~\mathcal{F}(\beta,\alpha)
\left[  \beta
,\alpha\right]  \ast\left(  \pi^{a}\left(\frac\lambda 2(\alpha+\beta) n\right)i\overleftrightarrow{\partial}%
_{+}~\pi^{b}\left(\frac\lambda 2(\alpha-\beta) n\right)\right)\, ,
\ee
We generalize  this operator to an arbitrary $N$ as
\be{Op:ON}
 O^{\left[  ab\right]  }(\lambda)=-~\mathcal{F}(\beta,\alpha)
\ast\text{AS~}\pi^{a}(({\scriptstyle\frac12}(\alpha+\beta)\lambda n)
~\overleftrightarrow{i\partial}%
_{+}~\pi^{b}({\scriptstyle\frac12}(\alpha-\beta)\lambda n)
\ee
where symbol "AS" denotes the antisymmetrization with respect to $O(N+1)$
indices:
\be{def:AS}
\text{AS~}\pi^{a}(x_{1})~\overleftrightarrow{\partial}_{+}~\pi^{b}%
(x_{2})=\frac{1}{2}\left[  \pi^{a}(x_{1})~\overleftrightarrow{\partial}%
_{+}~\pi^{b}(x_{2})-\pi^{b}(x_{1})~\overleftrightarrow{\partial}_{+}~\pi
^{a}(x_{2})\right]  .\
\ee
It is clear that the correspondence between the two definitions is given by
\be{ON2O3}
O^{c}(\lambda)=i\varepsilon\lbrack abc]\lim_{N\rightarrow 3}O^{\left[ ab\right]}.
\ee

Let us consider now the following matrix element in the extended theory
\be{1}
\int\frac{d\lambda}{2\pi}e^{-ip_{+}x\lambda}\left\langle \pi^{b^{\prime}%
}(p)\left\vert
 O^{[ab]}(\lambda)
\right\vert \pi^{a^{\prime}}(p)\right\rangle =4\left[  \delta^{aa^{\prime}%
}\delta^{bb^{\prime}}-\delta^{ab^{\prime}}\delta^{ba^{\prime}}\right]
~q(x).
\ee
It defines an analog of isovector PDF $q(x)$ for the $O(N+1)$ case,
the isovector pion GPD is defined by obvious generalization of the above equation.

The isoscalar PDF (and its generalization to GPD) for an arbitrary $N$ is defined as:

\be{2}
\int\frac{d\lambda}{2\pi}e^{-ip_{+}x\lambda}\left\langle \pi^{b}(p)\left\vert
 O(\lambda)
\right\vert \pi^{a}(p)\right\rangle =2 \delta^{ab}
~Q(x)\, ,
\ee
where the operator is defined as:
\be{3}
O(\lambda)=
-\mathcal{F}(\beta,\alpha)\ast\Big[\sigma(x_1\lambda)i\overleftrightarrow \partial_+\sigma(x_2\lambda)+
\pi^a(x_1\lambda)i\overleftrightarrow\partial_+ \pi^a
(x_2\lambda)\Big]
\ee
with
$$x_1~=~\frac{\alpha+\beta}{2}~,~~x_2~=~\frac{\alpha-\beta}{2}.$$

\begin{figure}[t]
\unitlength1mm
\begin{center}
\hspace{0cm}
\insertfig{9}{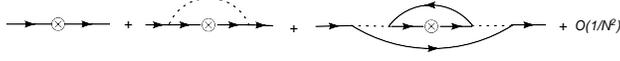}
\end{center}
\vspace{-0.5cm}
\caption[dummy]{  Diagrams which contribute to $1/N$ accuracy to the pion GPDs.
\label{fig1}}
\end{figure}

In order to understand the relation of the $1/N$ expansion to the standard $\chi$PT we have to inspect
 the diagrams of the large$-N$ approach. To the $1/N$ order for GPDs one has to compute the diagrams shown in Fig.\ref{fig1}.
The first diagram is the tree contribution to the pion PDF,
the second and the third diagrams are the first non-trivial $1/N$ corrections\footnote{To the $1/N$ order there are another tadpole-like diagrams, however
they do not contribute to the leading logs.}. The solid line
denotes the usual pion propagator and the dashed line is the
propagator of the auxiliary field $\varphi$ which is suppressed by $1/N$ (see details of $1/N$ expansion technique in Appendix~B).
The explicit expression for the propagator of the auxiliary field is given by
\be{dashed}
\left\langle \varphi\varphi\right\rangle &=&\frac{-1}{\Delta \FF}
\left[
1-\frac{N}{2 \Delta \FF}\,\, \insertfig{1}{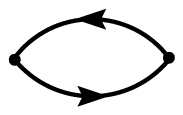}
\right]  ^{-1}, \quad \Delta=\frac{i}{p^2-m_\pi^2},\\
i
\begin{array}{c}
\insertfig{1}{loop}
\end{array}
&=&\frac{1}{\varepsilon}\frac{\mu^{2\varepsilon}}{[-p^2]^{\varepsilon}}
\frac{\Gamma(1+\varepsilon)\Gamma^2(1-\varepsilon)}{\Gamma(2-2\varepsilon)} \, ,
\ee
where we apply the dimensional regularization with $D=4-2\varepsilon$ and with renormalization scale $\mu$.
After expanding the above expression in powers of pion coupling
$N/\FF$:
\be{FFexp}
\left\langle \varphi\varphi\right\rangle   =\frac{-1}{\Delta \FF}\left(1+
\left(\frac{N}{2\Delta\FF}\right)^2\,\, \insertfig{1}{loop}
+\left(\frac{N}{2\Delta \FF}\right)^3\,\, \insertfig{1}{loop}\mskip-4mu\insertfig{1}{loop}
+\dots\right)\, ,
\ee
we observe that this expansion produces  the infinite series of the bubble chains with
the multiplicative factor proportional to $N/\FF$. This expansion explains which  diagrams of the usual
$\chi$PT are taken into account  at given order of $1/N$ expansion.

\section*{Large$-N$ resummation for the  pion PDFs }
In this section we consider the large-$N$ resummation of the
singular contributions to the pion PDFs. We give detailed
calculations for the case of the isovector PDF $q(x)$. For the
isoscalar case we outline main differences in the calculation
comparing to the isovector case.
\subsection*{Isovector PDF}
\begin{figure}[t]
\unitlength1mm
\begin{center}
\hspace{0cm}
\insertfig{6}{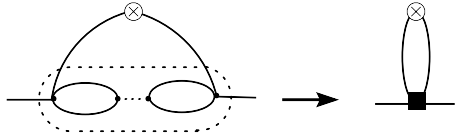}
\end{center}
\vspace{-0.5cm}
\caption[dummy]{   The diagram with the $n-$loop $UV-$counterterm. Corresponding counterterm is shouwn by the dashed line
\label{fig2}}
\end{figure}

Let us compute the leading logarithms
from the diagrams with the insertion of the bubbles and select the contributions with the derivatives of the $\delta-$functions.
Only two first diagrams in Fig.~\ref{fig1} contribute to the isovector PDF,
the third diagram  obviously contributes obviously to the isoscalar
operator only.
The corresponding diagrams are $UV-$divergent and must be renormalized. The $UV-$subtractions is performed
by the application of  {\bf R}-operation\footnote{An  introduction to the renormalization in quantum field theory
and detailed discussion of the {\bf R}-operation see in  \cite{ANV,Collins}.}. The series of the relevant diagrams originating from the loop diagram
in Fig.\ref{fig1} can be written as
\be{def:DN}
q(x) =\frac{1}{N}\sum_{n=0}^{\infty}%
{\rm \bf R}\, F^{I=1}(\beta,\alpha)\ast\int
\frac{d^D k}{(2\pi)^D}\, 
\frac{\mu^{2\varepsilon}\ k_+ \delta(\beta k_+ - x p_+) }
{\left[ k^{2}-m_\pi^{2}\right]  ^{2} }
\left(  \frac{N  }{\FF} [(k+p)^{2}- m_\pi^2] \right)^{n+1}
\left[
\insertfig{1}{loop}
\right]^{n}
\ee
The $\delta-$function in the numerator arises from the non-local vertex of the operator. It is clear that each term in the sum
\re{def:DN} corresponds to $n+1$-loop diagram.
The {\bf R}-operation removes the $UV$-divergencies in the $n+1$-loop diagram. The subtractions have following structure
\be{Rop}
{\rm \bf R}\ G^{(n+1)}  =  G^{(n+1)}
+\sum_{\{\Gamma_k \}}\frac{1}{\varepsilon^k} G^{(n+1-k)}
+\sum_{k=1}^n \frac1{\varepsilon^{k}} G^{(k)}_{\rm tree}\, ,
\ee
where $G^{(n+1)}$ denotes the bare $(n+1)$-loop diagram, the second term
describes the subtractions of the subdivergencies in the bare diagram,
the sum in the second term runs over the set of the $UV-$divergent
$k$-loop subdiagrams $\{\Gamma_k \}$.
The index $k$ denotes the number of the loops in the subdiagram and the pole $\frac{1}{\varepsilon^k}$ originates
from its $UV-$divergence. It is clear that the index $k$ takes values $1,\dots,n$.
The term $ G^{(n+1-k)}$ describes the $(n+1-k)$-loops diagram with insertion of
the counterterm vertex that results from the contraction of the $k$-loop subdiagrams.
The last term denotes the subtraction of the total resulting divergency,
$G_{\rm tree}^{(k)}$ is the notation for the tree level diagram with the
appropriate vertex structure. The contributions to the
coefficient in front of the leading logarithm arises from the terms in \re{Rop} except the last one.
 The $\varepsilon$ pole structure of the $(n+1-k)$-loop diagram $G^{(n+1-k)}$ in Eq.~\re{Rop} is the following:
\be{Gexp2}
\frac{1}{\varepsilon^k}  G^{(n+1-k)} &=&
\frac{ 1 }{\varepsilon^{n+1}}\ \left(\frac{\mu^2}{m_\pi^2}\right)^{\varepsilon(n+1-k)}G_{k}(\varepsilon)
\\
\nonumber
&=&\frac{1}{\varepsilon^{n+1}}
\left(
g^{(k)}_0+\varepsilon\ g^{(k)}_1\ln\left(\frac{\mu^2}{m_\pi^2}\right)+\dots
+\varepsilon^n g^{(k)}_n\ln^n\left(\frac{\mu^2}{m_\pi^2}\right)
\right )\\
\nonumber
&+&g^{(k)}_{0}\ \frac{(n+1-k)^{n+1}}{(n+1)!}\ \ln^{(n+1)}\left(\frac{\mu^2}{m_\pi^2}\right)+\dots .
\ee
In the above equation we took into account that $G_k(\varepsilon)$ is the polynomial of order $(n+1-k)$ in $\varepsilon$ and it is an analyitical function
of external momenta and the pion mass.
In the sum of all contributions in Eq.~\re{Rop} the poles in Eqs.~\re{Gexp2} with non-analytic dependence on external momenta and $m_\pi$
must cancel due to
locality of the total counterterm. That imposes  certain relations between the coefficients $g^{(k)}_{l}$ with $k=0,\dots,n$ and $l=0,\dots,n$.
One can easily show that these relations
leave only one unknown coefficient zhat must be computed in order to find the total coefficient in front of
$\ln^{(n+1)}(\mu^2/m_\pi^2)$, and has the form:
\be{Rop1}
{\rm \bf R}\ G^{(n+1)}  =\ln^{(n+1)}\left(\frac{\mu^2}{m_\pi^2}\right)\ \sum_{k=0}^n g^{(k)}_{0}\ \frac{(n+1-k)^{n+1}}{(n+1)!}\ +\dots \, .
\ee
It is convenient to introduce coefficient $g^{n}_{0}$ as an unknown quantity, i.e. one has to compute
only the contributions $\sum_{\{\Gamma_{n} \}}\frac{1}{\varepsilon^n} G^{(1)}$. In other words, we have to compute only one loop
diagram, but with all possible $n-$loop  counterterms. An example of  one of such diagrams is shown in Fig.~\ref{fig2}.
The structure of the $n-$bubble counterterm  has the form:
\be{bubc}
\frac{1}{\varepsilon^n}\left( \gamma_n (kp)^{n+1}+\left[ \gamma^{(1,0)}_n m_\pi^2+ \gamma^{(0,1)}_n k^2\right](kp)^{n}
+\dots
 \right)
\ee
where the dots denote all other possible  contributions constructed from the $m_\pi^2,\, k^2$ and $(kp)$.
The power $n$ is dictated by dimensions.
Inserting the expression \re{bubc} into the original diagram instead of the bubbles chain
one obtains
\be{Gred1}
 G^{(1)}\sim
F^{I=1}(\beta,\alpha)\ast\int
\frac{d^D k}{(2\pi)^D}\,
\frac{\mu^{2\varepsilon}\ k_+ \delta(\beta k_+ - x p_+) }
{\left[ k^{2}-m_\pi^{2} \right]^{2} }
\left( \gamma_n (kp)^{n+1}+\left[ \gamma^{(1,0)}_n m_\pi^2+ \gamma^{(0,1)}_n k^2\right](kp)^{n}
+\dots
 \right)
\ee
Now we observe that the most singular term (with the highest derivative of $\delta$-function) is produced from the contribution with
the highest power of the scalar product $(kp)$. Expanding the $\delta-$function in the numerator of Eq.~\re{Gred1}
\be{num}
k_+ \delta(\beta k_+ - x p_+)=k^{n+1}_+\,\beta^n \delta^{(n)}( x p_+ ) +\dots
\ee
and neglecting the terms with lower power of $k_+$, one obtains
\be{Gred2}
 G^{(1)}\sim
\gamma_n  \delta^{(n)}( x p_+ )\ [F^{I=1}(\beta,\alpha)\ast \beta^n ]\, \int
\frac{d^D k}{(2\pi)^D}\, \frac{\mu^{2\varepsilon}\ k^{n+1}_+ (kp)^{n+1} } {\left[
k^{2}-m_\pi^{2} \right]^{2} } \ee
Computing the one-loop integral  over $k$ we obtain the expression for $g_0^{(n)}=(n+1)!\ \gamma_n$. Using
Eq.~(\ref{Rop1}) we obtain
the desired singular contribution to PDF:
\be{Gred3}
q(x) \sim \gamma_n \delta^{(n)}( x )a^{n+1}_\chi \ln^{n+1}[1/a_\chi]+\dots\, ,
\ee
where  we fixed  the renormalization scale $\mu=(4\pi F_\pi)^2$ for simplicity.

\begin{figure}[t]
\unitlength1mm
\begin{center}
\hspace{0cm}
\insertfig{6}{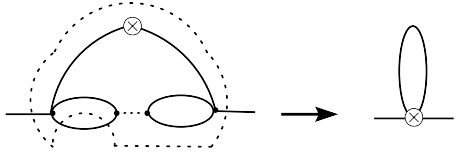}
\end{center}
\vspace{-0.5cm}
\caption[dummy]{  The diagram with the $n-$loop $UV-$counterterm which includes the operator vertex and resulting reduced one-loop
diagram. The counterterm is shown by the dashed line
\label{fig3}}
\end{figure}

Going back to the general case, we  compute the sum over all possible $n-$loop subdiagrams.
For that, in addition
to the considered case, one has to consider the subdiagrams with the operator vertex,
for instance such as in Fig.~\ref{fig3}.
Obviously, the reduced diagrams have always at least one contracted line, this reduces possible
power of derivatives of the $\delta-$functions in the resulting expression. Indeed, the contracted
propagator is always proportional
at least to one power of the pion
mass $m_\pi^2$ that excludes one power in possible expansion of the $\delta-$function. In other words,
there is no  possibility to produce the singular contributions of type \re{Gred3} from the contributions
with vertex subdiagrams in the diagrams with the bubble chains.

From the above consideration one learns several important lessons. We expect that each term in the sum of
diagrams \re{def:DN} contains a singular contribution proportional to $\delta^{(n)}( x )a^{n+1}_\chi \ln^{n+1}[1/a_\chi]$.
Moreover, taking into account the relations due to the cancellation of the poles with non-local dependence on external momenta,
we obtained that the calculations of these contributions can be reduced to a consideration of the diagram as in Fig.\ref{fig2}
 with $n-$loop bubble chain counterterm. Important technical observation, is that the structure of the corresponding counterterm
 is restricted by the maximal power of the scalar product $(kp)$ as in  \re{Gred1} and  does not depend on
 the pion mass in
 to the subdiagram. Therefore in the practical calculations we can put the pion mass inside the bubble chain to zero.
In addition, one can expand the $\delta-$function in expression \re{def:DN} and pick up only the contribution with
$\delta^{(n)}(x)$. Using these simplification and denoting the most singular term as $q^{\rm sing}(x)$ one obtains
\be{DNsimp}
q^{\rm sing}(x) =\frac{1}{N} \sum_{n=0}^{\infty}%
\langle x^n\rangle                      
\delta^{(n)}( x p_+)
{\rm \bf R}\,
\int
\frac{d^D k}{(2\pi)^D }\, 
\frac{k_+^{(n+1)}}
{\left[ k^{2}-m_\pi^{2}\right]  ^{2} }
\left(  \frac{N}{\FF} [(kp)^{2}] \right)^{n+1}
\left[
\insertfig{1}{loop}
\right]_{m_\pi=0}^{n}\, ,
\ee
where the {\bf R}-operation removes subdivergencies  only from the bubble chain subdiagram and
\be{def:mom}
\langle x^n\rangle=F^{I=1}(\beta,\alpha)\ast\beta^n=\int_{-1}^{1}dx x^n \chil{q}(x)\, .
\ee
 It turns out that such simplified
expressions can be computed directly, without the combinatorial analysis of the {\bf R}-operation mentioned above.
We provide the  corresponding technical details in the Appendix~B. The result is very simple
\be{DNres1}
q^{\rm sing}(x)=-\frac{2}{N} \sum_{n=0,2,4,\dots}^{\infty}%
 \delta^{(n)}( x ) \langle x^n\rangle
\frac{\epsilon ^{n+1}}{(n+1)!}~
\ee
where $\epsilon=\frac N 2 a_{\chi}\ln[1/a_\chi]$. This expression shows that one has the singular contributions  to all orders and we
can perform the summation of these terms. Let us represent the $n$-th derivative of the $\delta$-function as:
\be{dltn}
\delta^{(n)}( x )=\int \frac{d\lambda}{2\pi} (i\lambda)^n e^{i\lambda x}.
\ee
Changing the order of the integration the summation we obtain
\be{DNres2}
q^{\rm sing}(x)  =-\frac{\epsilon}{N}\int \frac{d\lambda}{2\pi}e^{i\lambda x}\int_{-1}^1 d\beta \chil{q}(\beta)
 \sum_{n=0,2,4,\dots}^{\infty}%
\frac{(i\lambda \beta \epsilon )^{n}}{(n+1)!}~
\ee
Computing the sum as
\be{evensum}
\sum_{n=0,2,4,\dots}^{\infty}%
\frac{(i\lambda \beta \epsilon )^{n}}{(n+1)!}~=\int_{-1}^1 d \tau \sum_{n=0}^{\infty}%
\frac{(i\lambda \beta \epsilon \tau )^{n}}{n!}
=\frac12 \int_{-1}^1 d \tau e^{i\lambda \beta \epsilon \tau}
\ee
and inserting it into \re{DNres2} we obtain
\be{DNres3}
q^{\rm sing}(x) =
-\frac{\epsilon}{N} \int_{-1}^1 d\beta \chil{q}(\beta)\int_{-1}^1 d \tau \delta(x+ \beta \epsilon \tau)
=-\frac{1}{N} \int_{-1}^1 d\beta \frac{\chil{q}(\beta)}{|\beta|}\theta(|x|/\epsilon<|\beta|).
\ee
Taking into account that  isovector PDF is symmetrical function $q(-x)=q(x)$ one can rewrite \re{DNres3} as
\be{DNres}
q^{\rm sing}(x) =-\frac{2\theta(|x|<\epsilon)}{N} \int_{|x|/\epsilon}^1 \frac{d\beta}{\beta} \chil{q}(\beta)
\ee
This is our final result for large-$N$ resummation of chiral corrections to the isovector pion PDF.
The above
equation demonstrates explicitly that resummation solves the problem of the singular terms -- the
infinite sum of $\delta$-function derivatives results in the contribution
of the type $f(x/a_\chi)/a_\chi$ as it was conjectured in Ref.~\cite{my}. Note that the large-$N$ resummation preserves all chiral properties of the considered quantities.

From \re{DNres},
one can observe that resummation does not change the small-$x$ asymptotic of the PDF. Taking the Regge-like
small-$x$ behaviour of PDF in the chiral limit:
\be{q0model}
\chil{q}(x)\sim \frac{1}{x^\omega}
\ee
we obtain non-analytic behavior of PDFs
in the chiral coupling $\sim (a_\chi\ln(1/a_\chi))^{\omega}$, for the chiral correction to PDF in
the small$-x$ region:
\be{smxas}
q^{\rm sing}(x)\sim \int_{x/\epsilon}^{1}\frac{d\beta}{\beta}\  \chil{q}(\beta)  \sim
\left(  \frac{ a_\chi\ln(1/a_\chi)}{ x }\right)^{\omega}\,.
\ee
Interestingly, the leading power of chiral coupling is determined by the intercept of
the Regge trajectory $\omega$ ($\omega \approx 1/2$ for the isovectors case).
This shows clearly importance of the resummation of singular chiral corrections for the
derivation of the leading chiral counting of PDF.
Naive chiral counting -- without taking into account the second scale related to the light-cone distance--suggests
that the leading chiral correction to the isovector PDF is $\sim a_\chi\ln(1/a_\chi)$.
However, it is not correct, as the leading
chiral corrections are dominated by the small-$x$ region
and one must perform the resummation discussed in the present paper.

\subsection*{Isoscalar PDF}
In the case of isoscalar ($C$ parity even) pion PDF one has to compute additionally the third diagram shown
in Fig.~\ref{fig1}.
The dashed line in this diagram represents the propagator of the auxiliary field (see Eq.~\ref{dashed}) corresponding
to the sum of bubble chains.
The third diagram is more complicated as it contains two loop integral. The expression for this diagram has the form:

\be{eq:domik}
&&\frac{1}{N}\sum_{n,m=0}^{\infty}
\Big(\frac{-N}{F^2_\pi}\Big)^{m+n+2}F(\beta,\alpha)\ast\int \frac{d^Dk}{(2\pi)^D}\frac{d^Dl}{(2\pi)^D}
\frac{k_+
\delta(k_+\beta-xp_+)[l^2-m_\pi^2]^{n+m+2}}{[k^2-m_\pi^2]^2[(l-p)^2-m_\pi^2][(l-k)^2-m_\pi^2]}\\
&\cdot& \frac{1}{\varepsilon^{m+n}}\frac{1}{[l^2]^{\varepsilon (m+n)}} \left(\frac{\Gamma(1+\varepsilon)\Gamma^2(1-\varepsilon)}{\Gamma(2-2\varepsilon)}\right)^{m+n}.
\ee
Applying the renormalization procedure ({\bf R} operation) and picking up the most singular in $x$ terms we obtain the contribution of the third diagram
to the singular part of the isoscalar PDF $Q^{\rm sing}(x)$:

\be{4}
Q^{\rm sing}_{\rm third}(x)=
\frac{-2}{N}\sum_{n,m=0}^{\infty} \langle x^{n+m+1}\rangle
\delta^{(m+n+1)}(x)\frac{(\frac N 2 a_\chi\ln(\frac{1}{a_\chi}))^{m+n+2}}{(m+n+3)!(n+m+1)}~,~~(m+n+1)=\text{odd}
\ee
Each term in the above sum depends only on $m+n$ combination of the summation indice; hence,
 this sum can be easily
computed. The final result of the summation of both the second and the third diagrams in Fig.~\ref{fig1} is the following:

\begin{equation}
Q^{\rm sing}(x)~=~
4\frac{\text{sign}(x)}{N}\theta(|x|<\epsilon)\int_{|x|/\epsilon}^1 \frac{d\beta}{\beta}\chil{Q}(\beta)
\Big(3-2\frac{|x|}{\epsilon\beta}\Big).
\end{equation}
Here, as before, $\epsilon~=~\frac N 2 a_\chi\ln(1/a_\chi)$.
As in the case of isovector PDF, the small $x$ behaviour of the chiral correction
has the same form as for the PDF in the chiral limit, e.g. the Regge -like that $\sim 1/x^\omega$.
One can easily calculate that the small-$x$ asymptotic of the leading chiral correction has the form:

\be{5}
Q^{\rm sing}(x) \sim \left(\frac{a_\chi\ln (1/a_\chi)}{x}\right)^\omega.
\ee
Again we see that the correct dependence of the small-$x$ PDF on the pion mass can be obtained
only after the resummation of all orders of the standard $\chi$PT.

Eq.~ (\ref{smxas}) and Eq.~(\ref{5}) show that the leading chiral dependence of the PDFs on $m_\pi$ is determined by the
small $x$ behaviour of the PDFs. The dynamics of the chiral degrees of freedom (Goldstone bosons)
and the dynamics of QCD degrees of freedom (quarks and gluons) intertwine. Noticeably, even the
power of the chiral expansion parameter changes with QCD evolution
of PDFs.

\section*{Large$-N$ resummation for the pion GPD }
\begin{figure}[t]
\unitlength1mm
\begin{center}
\hspace{0cm}
\insertfig{5}{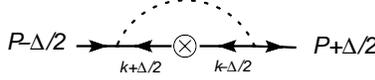}
\end{center}
\vspace{-0.5cm}
\caption[dummy]{
The diagram for $1/N$ correction for isovector GPD with  momentum flow.
\label{fig5}}
\end{figure}
In this section we consider the resummation as before but for the GPD matrix element.
The  diagrams describing the $1/N$ correction to the pion GPDs
$ H(x,\xi,t)$\footnote{We do not introduce special notation for this
correction
   in order to avoid proliferation of notations. It is clear from the context which part of $ H(x,\xi,t)$ is considered}
   are the same as in Fig.~\ref{fig1}, bu with momentum flow depicted in
   Fig.\ref{fig5}.
   The expression for the $1/N$ correction to the isovector GPD has the following form:
\be{def:GN}
H^{\rm I=1} (x,\xi,t)&=&\frac{1}{N}\sum_{n=0}^{\infty}%
{\rm \bf R}\, F^{I=1}(\beta,\alpha)\ast
\\
\nonumber
&&\int
\frac{d^D k}{(2\pi)^D}\, 
\frac{k_+ \delta(\beta k_+ +\alpha\xi - x p_+) }
{[(k-\Delta/2)^{2}-m_\pi^2][ (k+\Delta/2)^{2}-m_\pi^2] }
\left(  \frac{N  }{\FF} [(k+p)^{2}] \right)^{n+1}
\left[
\begin{array}{c}
\insertfig{1}{loop}
\end{array}
\right]^{n}\, .
\ee
 Due to the presence of the momentum transfer instead of
$\delta$-functions we obtain the  contributions $\sim \frac{\theta(|x|<\xi)}{\xi^{n+1}}$. Such contributions
in the forward limit ($\xi\to 0$) are reduced to the derivatives of $\delta-$function. Therefore we shall
separate them neglecting  the others less singular terms. The explicit resummation can be done in the same
way as for the PDF. But now the calculations are slightly more complicated due to additional parameters $\xi$ and $t$.
At the end we obtain the following expression for the singular contributions to the isovector pion GPD:
\be{barH1}
 H^{\rm I=1} (x,\xi,t)&=&
 -\frac{1}{N}~\frac{\theta(|x|\leq\xi)}{\xi^{n+1}}
\sum_{n=0,2,\dots}^{\infty} \int_{-1}^{1}d\eta
\frac{\left( \frac N 2 R\ln{1/R}~\right)  ^{n+1}}%
{(n+1)!}\frac{1}{(n+1)}
\\
\nonumber
& &  \partial_{\eta}^{n+1}
\eta F^{I=1}(\alpha,\beta)*\delta\left(\beta\eta+\alpha-\frac{x}{\xi}\right)
\ee
where
\[
R=\frac{m_{\pi}^{2}-t(1-\eta^{2})/4}{\left(  4\pi F_\pi\right)  ^{2}}\,
\]
\[
R|_{t=0}=a_\chi,~~ R|_{m_\pi=0}=(1-\eta^2)\ b_\chi,
\]
is the small chiral expansion parameter. One can easily perform summation of the singular contributions in Eq.~(\ref{barH1})
with the following result:
\be{dH1}
  H^{\rm I=1} (x,\xi,t) ={  -}\frac{1}{2{  N}}%
~\int_{-1}^{1}d\eta~\int_{-\zeta}^{\zeta}d\tau~~\left(   1+\frac{\eta\xi
}{\tau}\right)  F^{I=1}({  \beta,\alpha})\ast\delta\left(  \left[  \xi
\eta+\tau\right]  \beta+\xi\alpha-x\right)  ,
\ee
where
\[
\zeta=\frac{N}{2} R\ln(1/ R).
\]
 The result of the large -$N$ resummation of the singular contributions
to the isoscalar pion GPD has the form:

\be{dH0}
H^{I=0}({  x,\xi,t})  & =&{ -}\frac{1}{N}\int_{-1}
^{1}d\eta\int_{0}^{\zeta}d\tau~({ 3-2~\tau/\zeta})~~F^{I=0}(\beta
,\alpha)\ast\left\{  \left(  {  1+}\frac{\eta\xi}{\tau}\right)
\delta\left(  \left[  {  \eta\xi+\tau}\right]  {  \beta+\xi\alpha
-x}\right)  \right.  ~\\
\nonumber
& &\left.  -\left(  {  1-}\frac{\eta\xi}{\tau}\right)  \delta\left(
\left[  {  \eta\xi-\tau}\right]  {  \beta+\xi\alpha-x}\right)
-\frac{2\eta\xi}{\tau}\delta\left(  {  \eta\xi\beta+\xi\alpha-x}\right)
\right\}
\ee
Note that the small chiral expansion parameter $\epsilon=\frac N 2 R \ln(1/R)$ enters
these expressions always divided by $\xi$, meaning that
for $\xi\sim \epsilon$ the singular contributions are not suppressed by the small chiral parameter.
In next sections we shall analyze obtained expression for
resummed chiral corrections to GPDs and corresponding amplitude of hard exclusive processes on the pion target.

\subsection*{Parton distribution in the transverse plane}

To analyze the distribution of partons in the transverse plane one has to
take the limit  $\xi\rightarrow 0$ keeping $t\neq 0$ in GPDs.
One can easily perform this limit for the expressions
 (\ref{dH1},\ref{dH0}) with the results\footnote{In section we consider the case of
 $m_\pi=0$, because we focus our discuss on the dependence of GPD on $t$ and $\xi$. For the real world
 this case corresponds to the parametrically wide  domain
 of momentum transfer squared
 $m_\pi^2\ll -t \ll (4\pi F_\pi)^2$.}:
\be{dH1perp}
   H^{\rm I=1} (x,\xi=0,t=-\Delta^2_{\perp}) \equiv q^{\rm sing} (x,\Delta_{\perp})
  =-\frac{2}{N}\  \theta(|x|<\epsilon  )
\int_{\frac{x}{\epsilon}}^1
\frac{d\beta}{\beta}
\chil{q}(\beta)\ \sqrt{1-\frac{x}{\epsilon\beta}},
\ee
and
\be{dH0perp}
   H^{\rm I=0} (x,\xi=0,t=-\Delta^2_{\perp}) \equiv Q^{\rm sing} (x,\Delta_{\perp})
  &=&-\frac{4}{N}\  \theta(x<\epsilon )
\int_{\frac{x}{\epsilon}}^1 \frac{d\beta}{\beta} \chil{Q}(\beta)\\
\nonumber
&\times&
\Bigg[3\sqrt{1-\frac{x}{\epsilon\beta}}-\frac{x}{\epsilon \beta}
\ln\Bigg(\frac{1+\sqrt{1-\frac{x}{\epsilon\beta}}}{1-\sqrt{1-\frac{x}{\epsilon\beta}}}\Bigg)\Bigg],
\ee
where $\epsilon=\frac N 2 b_\chi \ln(1/b_\chi)$ with $b_\chi=|t|/4(4\pi F_\pi)^2=\Delta_{\perp}^2/4(4\pi F_\pi)^2$.

The Fourier transform of the PDFs $q(x,t),Q(x,t)$ in $\Delta_\perp$ ($\Delta_\perp^2=-t$) gives
the density of partons in transverse plane ($b_\perp$)  with given momentum fraction $x$ \cite{burk}.
The chiral expansion allows us to compute the asymptotics of the distribution in the transverse plane at $b_\perp \to \infty$.
The one-loop $\chi$PT gives the following large $b_\perp$ behaviour of isovector quark density $q(x,b_\perp)$\footnote{We remind that we consider the chiral limit
$m_\pi=0$.} \cite{kiv02}:
\be{bperp1}
q(x,b_\perp)\sim \frac{1}{b_\perp^4}\ \delta(x)\, .
\ee
The large $b_\perp$ behaviour of isoscalar distribution of quarks $Q(x,b_\perp)$ in two-loop  $\chi$PT has the form:

\be{bperp2}
Q(x,b_\perp)\sim \frac{\ln(b_\perp^2)}{b_\perp^6}\ \delta^\prime(x)\, .
\ee
We see that at large $b_\perp$ the density of (anti)quarks is formed by partons with small longitudinal momentum fraction $x$.
The presence of the $\delta$-functions is the artifact of the finite order $\chi$PT calculations, we simply can not resolve
the $\delta$-functions. This shows that we must perform the resummation of $\chi$PT in order to access the correct large $b_\perp$
asymptotic of
quark density in the transverse plane. We can easily perform the Fourier transormation of our results for $q(x,t)$
and $Q(x,t)$, see Eqs.~(\ref{dH1perp},\ref{dH0perp}). As we discussed above only partons with small $x$ are relevant for the large
$b_\perp$ asymptotic of parton densities. Assuming the Regge-like behaviour for the forward parton density $q(x)\sim 1/x^{\omega_-}$
and $Q(x)\sim 1/x^{\omega_+}$ (the subscript $\pm$
stays for the C-parity [signature] of the Regge trajectory) we obtain
the following result for $b_\perp\to\infty$:

\be{bperpasy}
&&q(x,b_\perp)\sim \frac{\ln^{\omega_-}(b_\perp^2)}{(b_\perp^2)^{1+\omega_-}}\ \frac{1}{x^{\omega_-}}\, ,\\
\nonumber
&&Q(x,b_\perp)\sim \frac{\ln^{\omega_+}(b_\perp^2)}{(b_\perp^2)^{1+\omega_+}}\ \frac{1}{x^{\omega_+}}\, .
\ee
The intercepts of the Regge trajectories are $\omega_-\approx 1/2$ and $\omega_+\approx 1.1$.
We see
that distribution of partons at large $b_\perp$ (\ref{bperpasy}) drops slower than the
naive results (\ref{bperp1},\ref{bperp2}) of finite order $\chi$PT.
Also we observe very interesting phenomenon --
the distributions of partons in the
transverse plane at large impact parameter $b_\perp$ depend on the intercept of the corresponding
Regge trajectory. This new phenomenon
is revealed due to the all order resummation of $\chi$PT developed in this paper.

The average transverse square size of a hadron in the transverse plane is defined as:

\be{size}
\langle b_\perp^2 \rangle(x)=\int d^2b_\perp\ b_\perp^2\ q(x,b_\perp)\, .
\ee
For the isoscalar PDF $Q(x,b_\perp)$ the integral for $\langle b_\perp^2 \rangle(x)$ is IR-finite in the chiral limit (see Eq.~\ref{bperp2}),
meaning that the corresponding radius is not determined by the chiral expansion. On contrary the integral for the isovector $\langle b_\perp^2 \rangle(x)$
is IR-divergent and, hence, its value is dominated by the large $b_\perp\sim 1/m_\pi$, so that we can obtain the $m_\pi$ behaviour of $\langle b_\perp^2 \rangle(x)$
in the isovector case:

\be{radius}
\langle b_\perp^2 \rangle(x)\sim \frac{1}{x^{\omega_-}}\ \frac{\ln^{\omega_-}(1/m_\pi^2)}{m_\pi^{2(1-\omega_-)}}.
\ee
We see again that at small $x$ the transverse size of the pion grows as the power\footnote{Note that in the Regge-like Ansatz for the $t$-dependence of GPDs the transverse size
of a hadron grows as $\sim \ln (1/x)$ at $x\to 0$.} of $1/x$, also its dependence on small pion mass is power like
(compare this with $\sim \ln(1/m_\pi^2)$ behaviour
of the isovector radius of the pion in the standard $\chi$PT). The growth of t
he nucleon transverse size at small $x$ has been discussed in Ref.~\cite{MS}.
The power-like
growth of the pion transverse size at small $x$
derived here implies that the similar phenomenon takes place for the nucleon case\footnote{We note that the results obtained
here for the isoscalar quark distribution are also valid for the gluon distributions.}.

Up to now we discussed the large $b_\perp$ asymptotic for the chiral limit $m_\pi=0$.
The case of $m_\pi\neq 0$ can be obtained from from the general result given in Eqs.~(\ref{dH1},\ref{dH0}):
\be{mpineq0}
q({  x,}b_\perp)\sim\frac{1}{N}\ \frac{1}{b_\perp^2}\ \left(  \frac{N}{|x|}\frac
{\ln\left[  b_\perp^{2} (4\pi F_\pi)^2 \right]}{b_\perp^{2}(4\pi F_\pi)^2}\right)
^{\omega}\left[  {  m_\pi b_\perp}\right]  ^{\omega+1}\int_{0}^{1}d\eta~(1-\eta^{2}%
)^{(\omega-1)/2}~K_{\omega+1}\left[  \frac{2{  m_\pi b_\perp}}{\sqrt{1-\eta^{2}}}\right]  .
\ee
For~$m_\pi b_\perp \rightarrow0:$
\[
q({  x,}b_\perp)\sim\frac{1}{N}\frac{1}{b_\perp^2}\left(  \frac{N}{|x|}\frac
{\ln\left[  b_\perp^{2} (4\pi F_\pi)^2 \right]}{b_\perp^{2}(4\pi F_\pi)^2}\right)
^{\omega}+\mathcal{O}(m_\pi b_\perp).
\]
This result coincides with the asymptotics (\ref{bperpasy}) for the massless pion.
For $ m_\pi b_\perp\rightarrow\infty$%
\[
q({  x,}b_\perp)\sim\frac{1}{N}\frac{e^{~-{  2m_\pi b_\perp}}}{b_\perp^{2}%
}\left(  \frac{N}{|x|}\frac{{  m_\pi b_\perp~}\ln\left[  b_\perp^{2}(4\pi F_\pi)^2%
\right]  }{b_\perp^{2}(4\pi F_\pi)^2}~\right)  ^{\omega}\left(1+\mathcal{O}\left(\frac{1}{m_\pi b_\perp}\right)\right)\, .
\]

\subsection*{Chiral corrections to the amplitudes of hard exclusive processes on the pion}
In this section we compute the contribution of the singular terms to the amplitude of
hard exclusive processes. We consider the Mandelstam $t$ in the region $m_\pi^2\ll -t\ll (4\pi F_\pi)^2$
in order to
focus on the the $t$-dependence of the amplitude, i.e. we can put effectively $m_\pi=0$.

The corresponding amplitude is computed in terms of GPDs as the following
convolution integral:

\be{defamplitude} {\cal
A}^{\rm I=0,1}(\xi,t)=\int_{-1}^{1}\frac{H^{\rm I=0,1}(x,\xi,t)}{\xi-x-i 0}\, .
\ee
Let us start with the isovector amplitude. To calculate the contribution of the singular terms
$\sim [b_\chi \ln(1/b_\chi)]^k/\xi k$
 into the amplitude
we substitute the expansion given by Eq.~(\ref{barH1}) into the integral (\ref{defamplitude}). The result
can be represented as the following series:

\be{ampbsum}
{\cal A}^{\rm I=1}(\xi,t)&=&-\frac{1}{N}%
\sum_{n=0,2,\dots}^{\infty}\left(  N b_{\chi}\ln{\left(\frac{1}{b_\chi}\right)}~\right)  ^{n+1}
\frac{1}{(n+1)}~\frac{1}{\xi^{n+1}}
 \int_{-1}^1 dz\ \frac{\chil{\Phi}^{\rm I=1}_{n+1}(z)}{1-z}\, .
\ee
Here we introduced the distribution amplitude (DA) of two pions with angular momentum $l=n+1$
defined as:
\be{6}
\chil{\Phi  }_{n+1}^{\rm I=1}(z)=\int_{-1}^{1}d\eta~ P_{n+1}(\eta)\chil{\Phi}^{\rm I=1}(z,\eta)\,,
\ee
where $P_{n+1}(\eta)$ are Legendre polynomials and the two pion DA \cite{Diehl} $\chil{\Phi}^{\rm I=1}(z,\eta)$
 is expressed in terms of coefficient function $F(\beta,\alpha)$ as follows:

\be{DA}
\chil{\Phi}^{\rm I=1}(z,\eta)=\eta F^{\rm I=1}(\alpha,\beta)*\delta(\beta\eta+\alpha-z)\,.
\ee

The sum (\ref{ampbsum}) has the form of the partial wave expansion of the amplitude
in the $t$-channel, see \cite{MVP99}.
Each term of the sum (\ref{ampbsum}) is real, however for $\xi \leq  N b_{\chi}\ln{(1/b_\chi)}$
the sum is divergent and the imaginary part of the amplitude is generated -- the case typical for
the duality. That naturally invites one
to apply the methods of the resummation of the $t$-channel exchanges developed for the dual
parametrization of GPD in Ref.~\cite{dual}. The final result for the chiral expansion of the imaginary part
of the amplitude is the following:
\be{im}
{\rm Im}{\cal A}^{\rm I=1}(\xi,t)&=&\pi\
\theta(\xi \leq 2 \epsilon) \int_{\xi/2\epsilon}^1
\frac{dx}{x}\ \chil{N}(x)\ \left(1-\sqrt{\frac{\xi}{2\epsilon x}} \right)\, ,
\\
{\rm Im} {\cal A}^{\rm I=0}(\xi,t)&=&\pi\ \theta(\xi \leq 2 \epsilon) \int_{\xi/2\epsilon}^1 \frac{dx}{x}\ \chil{N}(x)\ \sqrt{\frac{\xi}{2\epsilon x}}\
\left(5-\sqrt{\frac{\xi}{2\epsilon x}}\ \left(2+6\ \frac{\epsilon x}{\xi}\right) \right)\, .
\ee
Here $\epsilon=\frac N 2 b_\chi \ln(1/b_\chi)$ [$b_\chi =|t|/4(4\pi F_\pi)^2$],
$\chil{N}(x)$ is the so-called GPD quintessence function \cite{tomography,tamed}, which is expressed
via the imaginary part of the amplitude at $m_\pi^2=t=0$ with help of Abel tomography methods \cite{tomography,alena}:

\be{tomog}
\chil{N}(x)&=&\frac{2}{\pi}\ \frac{x(1-x^2)}{~~~(1+x^2)^{3/2}}
\int_\frac{2 x}{1+x^2}^1\frac{d\xi}{\xi^{3/2}}\
\frac{1}{\sqrt{\xi-\frac{2 x}{1+x^2}}} \left\{ \frac 12{\rm Im\ }
\chil{{\cal A}}(\xi)-\xi \frac{d}{d\xi}{\rm Im\ }\chil{{\cal A}}(\xi) \right\}\, ,\\
{\rm Im\ }\chil{{\cal A}}(\xi)&=&\pi\ \chil{H}(\xi,\xi)\, .
 \ee
Given the Regge-like small $\xi$ behaviour of the amplitude at $t=0$ --${\rm Im\ }\chil{{\cal A}}(\xi)\sim 1/\xi^\omega$. The leading chiral
correction to the amplitude has the form \cite{my}:

\be{7}
{\rm Im\ }{\cal A}(\xi,t) \sim \frac{1}{\xi^\omega}\ \left[b_\chi \ln\left(\frac{1}{b_\chi}\right)\right]^\omega\, ,
\ee
which depends on the intercept of the corresponding Regge trajectory ($\omega\approx 1/2$ for $I=1$ and
$\omega\approx 1.1$ for $I=0$).
This novel phenomenon can be obtained only after all order resummation of $\chi$PT.
The standard $\chi$PT gives the leading chiral correction of the form ${\rm Im\ }{\cal A}^{\rm I=1}(\xi,t) \sim \frac{1}{\xi^\omega}\left[b_\chi \ln\left(\frac{1}{b_\chi}\right)\right]$
and ${\rm Im\ }{\cal A}^{\rm I=0}(\xi,t) \sim \frac{1}{\xi^\omega}\left[b_\chi \ln\left(\frac{1}{b_\chi}\right)\right]^2$.

\section*{Conclusions and discussion}
We demonstrated that in the region of Bjorken $x_{\rm Bj}\sim m_\pi^2/(4\pi F_\pi)^2$
and/or $x_{\rm Bj}\sim |t|/(4\pi F_\pi)^2$
the ordinary chiral expansion for the pion PDFs and GPDs fails and one must perform all order resummation of the standard $\chi$PT.
We perform such resummation in the large-$N$ limit.
Explicit resummation allowed us to reveal novel phenomena
in quark mass expansion of PDFs, in low energy behaviour of GPDs and
amplitudes of hard exclusive processes.
The main qualitative (model independent) results are the following:

\begin{itemize}
\item
The leading small $m_\pi$  asymptotic in the region of small $x$ of pion PDFs depends on the intercept ($\omega$) of the corresponding Regge trajectory\footnote{We consider
the Regge-like behaviour of PDFs for simplicity, one can easily obtain corresponding leading chiral corrections for other types of small $x$ behiour of PDFs.}:
\be{a}
q(x)\sim \frac{1}{x^\omega}\ \left[\frac{m_\pi^2}{(4\pi F_\pi)^2} \ln\left(\frac{1}{m_\pi^2}\right)\right]^\omega
\ee
\item
The leading large impact parameter ($b_\perp$) asymptotics of the quark distribution in the transverse plane has the form (we show result for $m_\pi$,
the result for $m_\pi\neq 0$ is given in Eq.~(\ref{mpineq0})):
\be{8}
q(x,b_\perp)\sim \frac{1}{x^\omega}\  \frac{\ln^{\omega}(b_\perp^2)}{(b_\perp^2)^{1+\omega}}\,.
\ee
The distribution of quarks at large impact parameter is controlled completely by the all order resummed
$\chi$PT developed in this paper. This asymptotic
is determined by the small-$x$ behaviour of usual PDFs, hence
this asymptotic depends on the scale, at which the corresponding PDF is defined. This is new and interesting
result -- the chiral expansion meets the QCD evolution.

\item
The leading small $t$ behaviour of the amplitude for hard exclusive processes on the pion target has the form:

\be{b}
{\rm Im\ }{\cal A}(\xi,t) \sim \frac{1}{\xi^\omega}\ \left[\frac{|t|}{4(4\pi F_\pi)^2} \ln\left(\frac{1}{|t|}\right)\right]^\omega\, .
\ee
Measurements of such processes at small $x_{\rm Bj}$ and small $t$ would allow us to probe the
chiral dynamics in a completely new regime --
the dynamics of chiral and  quark--gluon degrees of freedom intertwines.
\end{itemize}
The complete results for the resummation of $\chi$PT for pion PDFs and GPDs are given in the main body of the paper.

In order to perform the resummation of $\chi$PT we used $1/N$ expansion. To assess the accuracy of the $1/N$ expansion we computed the first three
coefficients in Eq.~(\ref{QchptD}) for arbitrary $N$ with the result:

\be{c}
D_1=-1,~~ D_2=-\frac 56\ N \left(1-\frac 1 N \right)\langle x\rangle, ~~ D_3=-\frac{1}{24}\ N^2 \left(1-\frac{37}{18}\ \frac 1 N+\frac{49}{18}\ \frac{1}{N^2} \right)\langle x^2\rangle\, .
\ee
For $N=3$ the above result coincides with the three loop calculation (see Eq.~(\ref{3loops})) of Ref.~\cite{my}, also for $N=1$ the $D_2$ coefficient is zero as it should be, because
the $O(2)$ $\sigma$-model is a free field theory. All that provides strong check of the three loop calculations of Ref.~\cite{my}.
From Eq.~(\ref{c}) we see that the $1/N$ corrections are sizable (30-40 \%) and seem to
increase with number of loops. To check this tendency we computed four and six loop
$D_4$ and $D_6$ coefficients in the expansion (\ref{QchptD}) with the results:)

\be{d2n}
\nonumber
D_4&=&-\frac{7 N^3}{480}\left(
1-\frac{76}{27}\frac{1}{N}+\frac{113}{27} \frac{1}{N^2}-\frac{64}{27}
   \frac{1}{N^3}\right)\langle x^3\rangle,\\
\nonumber
D_6&=&-\frac{N^5}{8960}\left(1-\frac{46208}{10125}\frac{1}{N}+\frac{9441001}{911250}\frac{1}{N^2}-\frac{36581227}{2733750}\frac{1}{N^3}+\frac{30246239}{2733750}\frac{1}{N^4}-\frac{2449121
}{546750}\frac{1}{N^5}\right)\langle x^5\rangle\, .
\ee
These coefficients are zero at $N=1$, that provides a check of our calculations. We see that indeed the $1/N$ corrections increase with number of loops (they amount to $60-75\%$).
However, we note that the total coefficients $D_n$ at $N=3$ decrease rapidly with number of loops, that indicates that the resummation of singular contributions
is possible beyond the large-$N$ expansion and that one expects that main qualitative conclusions of the present paper are valid also for the exact resummation.

\section*{Acknowledgments}

This work was supported in parts by the Alexander von Humboldt Foundation, by BMBF,
 by the Deutsche Forschungsgemeinschaft,
 the Heisenberg--Landau Programme grant 2007,
 and the Russian Foundation for Fundamental Research
 grants No.\ 06-02-16215 and 07-02-91557
\bigskip

\section*{Appendix~A. Twist-2 operators and their matrix elements}
In this Appendix we briefly describe the  definitions and some technical
details used in the paper. We introduce two light-like vectors $n$ and $ \bar n$:
\be{llv}
n^2=\bar n^2=0,  n\cdot\bar n=1,\,\, a_{+}=a\cdot n.
\ee
There exist two QCD quark light-cone operators of twist-2:%
\be{PLR}
P_{R}=\frac{1}{2}\left(  1-\gamma_{5}\right)  ,~~P_{L}=\frac{1}{2}\left(
1+\gamma_{5}\right)  ,
\ee
\be{OLR}
\left[  O_{R}\right]  _{fg}  & =\bar{q}_{g}\left(  \frac{1}{2}\lambda
~n\right)  ~\gamma_{+}P_{R}~q_{f}\left(  -\frac{1}{2}\lambda~n\right)  ,\\
\left[  O_{L}\right]  _{fg}  & =\bar{q}_{g}\left(  \frac{1}{2}\lambda
~n\right)  ~\gamma_{+}P_{L}~q_{f}\left(  -\frac{1}{2}\lambda~n\right)  ,
\ee
where indexes $f,g$ stand for flavor. These operators transform
under the global chiral rotations  as
\be{Chrot}
  O_{L}\rightarrow V_L   O_{L} V_L^\dag
\,,\,\,
 O_{R}\rightarrow V_R   O_{R} V_R^\dag\, .
\ee
In $\chi$PT these QCD operators are described by an effective chiral operator with
unknown chiral constants. In the pure pion sector
one finds \ci{kiv02,man,che}
\be{matchingL}
O^L_{fg}(\lambda)&=&
-\frac{i F_\pi^2}{4}\
\mathcal{F}(\beta , \alpha)*
 \left[U\left(\frac{\alpha+\beta}{2}\lambda n\right)n\cdot\partialboth
U^\dagger\left(\frac{\alpha-\beta}{2}\lambda n\right)\right]_{fg} \, ,
\\
O^R_{fg}(\lambda)&=&
-\frac{i F_\pi^2}{4}\  \mathcal{F}(\beta , \alpha)*
 \left[U^\dagger\left(\frac{\alpha+\beta}{2}\lambda n\right)n\cdot\partialboth
U\left(\frac{\alpha-\beta}{2}\lambda n\right)\right]_{fg}\, .
\label{matchingR}
\ee
where by asterisk we denote the integral convolution with
respect to $\beta$ and $\alpha$:
\be{asterisk}
\mathcal{F}(\beta , \alpha)*O(\beta,\alpha)
\equiv
 \int_{-1}^1 d\beta \int_{-1+|\beta|}^{1-|\beta|} d\alpha\
\mathcal{F}(\beta , \alpha)\, O(\beta,\alpha)\, .
\ee
Here $\mathcal{F}(\beta , \alpha)$ represents the real generating function
for the tower of
low-energy constants and $\partialboth_\mu$ denotes a combination of derivatives
$\stackrel{\rightarrow}{\partial_\mu}-\stackrel{\leftarrow}{\partial_\mu}$.
It is important to note that the expressions
 \re{matchingL} and \re{matchingR} describe correctly only the operator
 vertices with two attached pions (including the tadpoles loops).
The low-energy constants $\mathcal{F}(\beta , \alpha)$ characterizes
the structure of the pion, they are not determined in the effective
field theory. According to the isospin $I=0,1$ one can construct two
independent functions:%
\be{def:FI}
F^{I=0}\left[  \beta,\alpha\right]   & =\frac12(\mathcal{F}\left[  -\beta
,\alpha\right]  -\mathcal{F}\left[  \beta,\alpha\right])  ,\\
F^{I=1}\left[  \beta,\alpha\right]   & =\frac12(\mathcal{F}\left[  -\beta
,\alpha\right]  +\mathcal{F}\left[  \beta,\alpha\right] ) ,
\ee
which are convenient for description of the pion matrix elements. Pion PDFs are
defined as
\be{def:PDFs}
\int\frac{d\lambda}{2\pi}e^{-ip_+\,x\lambda}\left\langle \pi^{b}(p^{
})\left\vert \text{tr}\left[  \tau^{c}O_{L+R}(\lambda)\right]  \right\vert
\pi^{a}(p)\right\rangle  & =4i\varepsilon\lbrack abc]q(x),\\
\int\frac{d\lambda}{2\pi}e^{-ip_+\, x\lambda}\left\langle \pi^{b}(p)
\left\vert \text{tr}\left[  O_{L+R}(\lambda)\right]  \right\vert \pi
^{a}(p)\right\rangle  & =2\delta^{ab}Q(x)\, ,
\ee
which in the chiral limit can be written as
\be{9}
2 \int_{-1+\left\vert \beta\right\vert }^{1-\left\vert \beta\right\vert }%
d\alpha~F^{I=0}(\beta,\alpha)~  & =\left[  \theta(\beta
)~\chil{q}(\beta)-\theta(-\beta)~\chil{\bar{q}}(-\beta)\right]  =\chil{Q}(\beta),\\
\int_{-1+\left\vert \beta\right\vert }^{1-\left\vert \beta\right\vert }%
d\alpha~F^{I=1}(\beta,\alpha)~  & =\theta(\beta)~\chil{q}(\beta)+\theta(-\beta
)~\chil{\bar{q}}(-\beta)=\chil{q}(\beta).
\ee
The GPDs are defined as:
\be{def:GPDs}
\int\frac{d\lambda}{2\pi}e^{-iP_+\, x\lambda}\left\langle \pi^{b}(p^{\prime
})\left\vert \text{tr}\left[  \tau^{c}O_{L+R}(\lambda)\right]  \right\vert
\pi^{a}(p)\right\rangle  & =4i\varepsilon\lbrack abc]H^{I=1}(x,\xi,t),\\
\int\frac{d\lambda}{2\pi}e^{-iP_+\,x\lambda}\left\langle \pi^{b}(p^{\prime
})\left\vert \text{tr}\left[  O_{L+R}(\lambda)\right]  \right\vert \pi
^{a}(p)\right\rangle  & =2\delta^{ab}H^{I=0}(x,\xi,t)
\ee
with~$P=\frac{1}{2}(p+p^{\prime}),~\xi=-\frac{(p^{\prime}-p)_+}
{(p^{\prime}+p)_+},~t=(p^{\prime}-p)^{2}.~$
In the forward limit $\xi\rightarrow0,\, t\rightarrow 0 $:%
\be{fwd}
H^{I=1}(x,0,0)=q(x),~\ H^{I=0}(x,0,0)=Q(x).
\ee

\section*{Appendix B. Large $N$ expansion in $O(N+1)$ model}
In this Appendix we discuss the technique of $1/N$ expansion for the $O(N+1)$ sigma model. We follow closely
the methods of $1/N$ expansion presented in the book by A.N.~Vasiliev \cite{ANV}.

\subsection*{Diagram technique for the  large$-N$ expansion}

Our task is to construct the large$-N$ expansion for the field theory given by
the Lagrangian
\begin{align}
\mathcal{L}  &  =-\frac{1}{2}\pi^{a}\partial^{2}\pi^{a}+N~V(\pi^{2}%
/N)+J^{a}\pi^{a},~\ \ \ \\
V(\pi^{2}/N)  &  =\frac{1}{8}\frac{\left(  \partial_{\mu}\pi^{2}/N\right)
^{2}}{\left(  G^{2}-\pi^{2}/N\right)  }+m^{2}G\sqrt{G^{2}-\pi^{2}%
/N}~,~\ \ G^{2}=F^{2}/N\sim\mathcal{O}(N^{0})
\end{align}
where $J^a$ is a source for the pion field. The constants $m$ and $F$ represent the bare values of the pion mass and
axial coupling.

Introducing two auxiliary fields $\psi$ and $\varphi$ and using the
representation
\[
1=\int D\psi~\delta(\pi^{2}-N~\psi)=\int D\psi~D\varphi\exp\left[
i~\varphi\left(  N~\psi-\pi^{2}\right)  /2\right]
\]
one can rewrite the Lagrangian as
\[
\mathcal{L}=-\frac{1}{2}\pi^{a}\partial^{2}\pi^{a}+\varphi\left(  N~\psi
-\pi^{2}\right)  /2+NV(\psi)+J^{a}\pi^{a},
\]
Performing the integration with respect to the pion fields we obtain the effective
Lagrangian
\[
\mathcal{L}_{eff}=N\frac{i}{2}\text{tr}\ln\left[  \left(  \partial^{2}
+\varphi\right)  /\partial^{2}\right]  +N~\varphi\psi/2+NV(\psi)+J^{a}
(\partial^{2}+\varphi)^{-1}~J^{a}/2~.
\]
Notice that the second operation $\partial^{2}$ arises from the normalization
constant in the functional integral. Computing the resulting integral by the steepest
descent method, one obtains the desired large$-N$ expansion.

The equations of motion define the constant solutions $\varphi=\varphi_{0}$
and $\psi=\psi_{0}:$%
\begin{equation}
~\varphi_{0}=\frac{m^{2}G}{\sqrt{G^{2}-\psi_{0}}},~~\psi_{0}=\int\frac{d^{4}%
k}{(2\pi)^{4}}\frac{i}{k^{2}-m^{2}}. \label{EOM}%
\end{equation}
Let us postpone the discussion of the solution and suppose simply that it
exists. Then performing expansion around the classical values we obtain the
effective action
\begin{align}
\mathcal{L}_{eff}  &  =N\frac{i}{2}\text{tr}\ln\left[  (\partial^{2}%
+\varphi_{0}+\varphi)/\partial^{2}\right]  +N~\left(  \varphi+\varphi
_{0}\right)  \left(  \psi+\psi_{0}\right)  /2\label{Leff1}\\
&  ~\ \ \ \ \ \ +NV(\psi+\psi_{0})+J^{a}\left(  \partial^{2}+\varphi
_{0}+\varphi\right)  ^{-1}~J^{a}/2~
\end{align}%
\begin{align}
&  =-\frac{N}{4}\varphi(x)i\Delta^{2}(x,x)\varphi(x)+N~\varphi(x)\psi
(x)/2-\frac{N}{2}\psi K_{\psi}\psi\label{Leff2}\\
&
~\ \ \ \ \ \ \ \ \ \ \ \ \ \ \ \ \ \ \ \ \ \ \ \ \ \ \ \ \ \ \ \ \ \ \ -J^{a}%
(x)\Delta(x,y)\varphi(y)\Delta(y,z)~J^{a}(z)/2+~...
\end{align}
where we introduced convenient notations:
\[
K_{\psi}=\frac{1}{4}\left(  G^{2}-\psi_{0}\right)  ^{-1}\, \partial^{2},
~~(\partial^{2}+\varphi_{0})^{-1}=i~\Delta(x,y),
\]
In the second line (\ref{Leff2}) we show the relevant to our accuracy
$\mathcal{O}(N^{-1})$ contributions only . The first term in (\ref{Leff2}) arises
from the trace in the first line, and $\Delta^{2}(x,x)$ in graphical notation
corresponds to the scalar loop
\begin{equation}
\Delta^{2}(x,x)\equiv L=\int\frac{d^{D}l}{(2\pi)^{D}}\frac{i}{l^{2}%
-\varphi_{0}}\frac{i}{(l+p)^{2}-\varphi_{0}}. \label{Ldef}%
\end{equation}
The quadratic in $\psi$ term originates from the expansion of the
interaction $NV(\psi)$ and the dots denote the terms that contribute to the
higher order corrections $\sim\mathcal{O}(N^{-2})$. We also keep the
contribution quadratical with respect to the sources because we consider only
the two-pion matrix elements. Performing the diagonalization with the help of
substitution $\psi\rightarrow\psi+\frac{1}{2}K_{\psi}^{-1}\varphi$ we obtain
a simple expression for the relevant part of the effective action
\begin{equation}
\mathcal{L}_{eff}=-\frac{N}{2}\varphi\left[  \frac{i}{2}~L-\frac{\left(
G^{2}-\psi_{0}\right)  }{\partial^{2}}\right]\varphi
-J^{a}\Delta\varphi\Delta~J^{a}/2~+~...~\ .
\end{equation}
The quadratical term provides the propagator for the auxiliary field $\varphi
$. Notice that insertion of this propagator produces the factor $1/N$. This
provides the counting rules with respect to $N$. The diagrams in Fig.\ref{fig1} are
produced when we contract all the auxiliary fields, differentiate with respect
to sources and amputate obtained irreducible diagrams. One can easily check
that the terms denoted as dots can provide only graphs which are suppressed as
$1/N^{2}$.

Let us shortly discuss the solutions of the equations (\ref{EOM}). It is clear
that constant $\varphi_{0}$ provides the physical mass $m_{\pi}$ of the pion
in the large$-N$ expansion and the second constant $\psi_{0}$ can be
associated with the physical axial coupling $F_{\pi}$.
 On can easily see that all diagrams which contribute
to the renormalization of the mass and coupling can not provide leading log's
to the singular terms because renormalization decreases the power of the
logarithm in front of $\delta-$function. Therefore we can skip the discussion
about the renormalization of the pion mass $m$ and axial coupling $F$.
Moreover, from the consideration in the text, we know that in the four-pion
blob we can neglect the mass of the pion because corresponding terms
produce subleading logarithms only. Therefore, in the propagator of the
auxiliary field $\varphi$ one can neglect the pion mass. That is equivalent to%
\begin{equation}
\varphi_{0}=m_{\pi}^{2}\rightarrow0,
\end{equation}
It is clear that we must keep the pion mass only in the propagators connecting the
operator vertex with the four-pion subdiagram. Next, in the $\varphi
-$propagator we substitute
\begin{equation}
\left(  G^{2}-\psi_{0}\right)  =F_{\pi}^{2}/N\equiv G_{\pi}^{2}%
\end{equation}
Finally, we obtain the following simple Lagrangian%
\begin{equation}
\mathcal{L}_{eff}\simeq-\frac{N}{2}\varphi\left[  \frac{i}{2}~L_{0}%
-\frac{G_{\pi}^{2}}{\partial^{2}}\right]  \varphi-J^{a}\Delta\varphi
\Delta~J^{a}/2~\ ,
\end{equation}
where $L_{0}$ denotes the massless scalar loop (\ref{Ldef}). Thus for the
propagator of the auxiliary field one has%
\begin{equation}
\left\langle \varphi\varphi\right\rangle =\frac{i}{N}\frac{p^{2}}{G_{\pi}^{2}%
}\frac{1}{1+\frac{p^{2}}{G_{\pi}^{2}}\frac{i}{2}L_{0}}. \label{prop}%
\end{equation}

\subsection*{Calculation of\ Large$-N$ corrections to PDF}

Our aim is to discuss the evaluation of $1/N-$corrections given by graphs in
Fig.\ref{fig1}. Consider one loop diagram which contributes to $q^{\rm sing}$ in
(\ref{DNsimp}).\ Expanding the propagator (\ref{prop}) in series with respect to loop
$L_{0}$ we obtain expansion of $q^{\rm sing}(x)$ which can be written as
\begin{equation}
q^{\rm sing}(x)=-\frac{2}{N}\sum_{n=0}^{\infty}\left\langle x^{n}\right\rangle
\delta^{(n)}(x)~J_{n}.\label{qbar}%
\end{equation}
Hence, we have to work with the expression which include insertion of the
products of the $n$ massless loops $\left(  \frac{p^{2}}{G_{\pi}^{2}}\frac
{i}{2}L_{0}\right)  ^{n}$ in each $J_{n}$. It is clear that such construction
is associated with the $n+1-$loop graph. The renormalization of such graph
includes the substruction of various subdivergencies. To the leading
logarithmic accuracy it is enough to consider the insertion of the
$L_{0}-$counterterms  only. The explicit expression for $L_{0}$ reads
\begin{equation}
\frac{N}{F_{\pi}^{2}}\frac{ip^{2}}{2}L_{0}=\frac{p^{2}}{F_{\pi}^{2}}\frac
{N}{2}\frac{1}{(4\pi)^{2}}\left(  \frac{\mu^{2}}{-p^{2}}\right)
^{\varepsilon}\frac{\Pi(\varepsilon)}{\varepsilon}=\frac{p^{2}}{m_{\pi}^{2}}%
\frac{\epsilon}{\varepsilon}+\mathcal{O}(\varepsilon^{0}),~\label{L0}%
\end{equation}
where for simplicity we introduced convenient notation%
\[
\epsilon=\frac{N}{2}\frac{m_{\pi}^{2}}{(4\pi F_{\pi})^{2}},~\ \Pi(\varepsilon
)=(4\pi)^{\varepsilon}\frac{\Gamma(1+\varepsilon)\Gamma(1-\varepsilon)^2}{\Gamma(2-2\varepsilon)}\
\]
For the product of $n~$loops $L_{0}$ we obtain obviously:%
\begin{equation}
\left[  \frac{p^{2}}{G_{\pi}^{2}}\frac{i}{2}L_{0}\right]  ^{n}=\left[
~\frac{p^{2}}{m_{\pi}^{2}}\left(  \frac{\mu^{2}}{-p^{2}}\right)  ^{\varepsilon}%
\frac{\epsilon~\Pi(\varepsilon)}{\varepsilon}\right]  ^{n}%
\end{equation}
The expression for the total unrenormalized graph $  J^B_{n}$
with $n$ loops $L_{0}$ can be represented as%
\begin{equation}
J_{n}^{B}=~\epsilon^{n+1}\frac{\Pi^{n}}{\varepsilon^{n}}\int\frac{d^{D}k}%
{\pi^{D/2}}\frac{i\left(  -k_{+}/p_{+}\right)  ^{\left(  n+1\right)  }%
}{\left[  {   -k}^{2}{   +m}_\pi^{2} \right]  ^{2}\left[  {   -(p+k)}%
^{2}{   /\mu}^{2}\right]  ^{(\Delta-\varepsilon)}}\left[  \frac
{2{   (pk)}}{{   m}_\pi^{2}}\right]  ^{(n+1)}~\text{~},\label{Jn}%
\end{equation}
where $\Delta\equiv\left(  n+1\right)  \varepsilon$. Evaluation of this
integral is standard. Rewriting the denominator with the help of Feynman
parameters, shifting the integral momentum $k$ and evaluating the resulting
integral we obtained%

\begin{align}
J_{n}^{B}(\varepsilon,\Delta)  & =\epsilon^{n+1}\frac{\Pi^{n}}{\varepsilon
^{n}}\frac{1}{{   \Delta}}\frac{\left(  \mu^{2}/m_\pi^{2}\right)  ^{\Delta}%
}{n!({   n+2})}\frac{~~\Gamma\left[  {   1-\Delta,1+\Delta
,D/2+n+1,3n+5-2}\Delta\right]  \ }{\Gamma\left[  { n+2-\Delta
,~D/2,3n+5-}\Delta-\varepsilon\right]  }\\
& \equiv\frac{\epsilon^{n+1}}{\varepsilon^{n+1}}\frac{1}{({   n+1)}%
}\left(  \frac{\mu^{2}}{m_\pi^{2}}\right)  ^{\Delta}~E_{n}\left[  \varepsilon
,\Delta\right]  .\label{JnB}%
\end{align}

Subtraction of $L_{0}-$subdivergencies in the $n-$loop chain $\left[
L_{0}\right]  ^{n}~$ is described by the substitution (\ref{L0}):
\begin{equation}
\left(  \frac{N}{F_{\pi}^{2}}\frac{ip^{2}}{2}L_{0}-\frac{p^{2}}{m_\pi^{2}}%
\frac{\epsilon}{\varepsilon}\right)  ^{n}=\sum_{k=0}^{n}\left(
\begin{array}
[c]{c}%
n\\
k
\end{array}
\right)  \left(  \frac{N}{F_{\pi}^{2}}\frac{ip^{2}}{2}L_{0}\right)
^{n-k}\left(  -\frac{p^{2}}{m_\pi^{2}}\frac{\epsilon}{\varepsilon}\right)  ^{k}%
\end{equation}%
\[
=\sum_{k=0}^{n}\left(
\begin{array}
[c]{c}%
n\\
k
\end{array}
\right)  \frac{(-\epsilon)^{n+1}}{\varepsilon^{k}}~~\left(  \frac{p^{2}}%
{m_\pi^{2}}\right)  ^{n}\left[  \Pi\left(  \frac{\mu^{2}}{-p^{2}}\right)
^{\varepsilon}\right]  ^{n-k}.~
\]
Each term in the last sum can be represented in the form (\ref{Jn}) with
$\Delta\rightarrow\Delta^{\prime}\equiv(n+1-k)\varepsilon$. Hence after
subtraction of the subdivergencies we obtain  following expression
\begin{equation}
J_{n}=~\sum_{k=0}^{n}\left(
\begin{array}
[c]{c}%
n\\
k
\end{array}
\right)  \frac{(-1)^{k}}{\varepsilon^{k}}~J_{n}^{B}(\varepsilon,\Delta
^{\prime})
\end{equation}
Substituting the $J-$integral as in (\ref{JnB}) we obtain%
\begin{equation}
J_{n}=~\frac{\epsilon^{n+1}}{\varepsilon^{n+1}}~\sum_{k=0}^{n}\left(
\begin{array}
[c]{c}%
n\\
k
\end{array}
\right)  \left(  \frac{\mu^{2}}{m_\pi^{2}}\right)  ^{\Delta^{\prime}}\frac{\left(
-1\right)  ^{k}~}{(n-k+1)}\text{~}~E_{n}\left[  \varepsilon,\Delta^{\prime
}\right]  .\label{RLGn}%
\end{equation}
Performing expansion of this expression with respect to $\varepsilon$ we
obtain the local poles and finite terms:
\begin{equation}
J_{n}=\frac{Z_{n+1}}{\varepsilon^{n+1}}+...+\frac{Z_{1}}{\varepsilon}%
+J_{n}^{\text{LL}}\ln^{n+1}\left[  \mu^{2}/m^{2}\right]  +\mathcal{O}(\ln^{n}\left[
\mu^{2}/m^{2}\right]  )
\end{equation}
In order to compute coefficient $J_{n}^{\text{LL}}$ we take into account that the largest
power $\ln^{n+1}$ arises only from the expansion%
\[
\frac{1}{\varepsilon^{n+1}}\left(  \frac{\mu^{2}}{m_\pi^{2}}\right)
^{\Delta^{\prime}}=\frac{(n-k+1)^{n+1}}{(n+1)!}\ln^{n+1}\left[  \mu^{2}%
/m_\pi^{2}\right]  +\mathcal{O}(\ln^{n}\left[  \mu^{2}/m_\pi^{2}\right]  )
\]
Picking up coefficient in front of $\ln^{n+1}$ \ in (\ref{RLGn}) we obtain:
\begin{align*}
J_{n}^{\text{LL}} &  =\left(  \epsilon\ln\left[  \mu^{2}/m_\pi^{2}\right]
\right)  ^{n+1}\frac{E_{n}\left[  0,0\right]  }{(n+1)!}\sum_{k=0}^{n}\left(
\begin{array}
[c]{c}%
n\\
k
\end{array}
\right)  \left(  -1\right)  ^{k}~(n-k+1)^{n}~\\
&  =\left\langle x^{n}\right\rangle \left(  \epsilon\ln\left[  \mu^{2}%
/m_\pi^{2}\right]  \right)  ^{n+1}\frac{n!}{(n+1)!},
\end{align*}

where$~~\ $we used that%
\begin{equation}
\sum_{k=0}^{n}(-1)^{k}(n+1-k)^{n}\left(
\begin{array}
[c]{c}%
n\\
k
\end{array}
\right)  =n!~,~~\lim_{\varepsilon\rightarrow0}E_{n}\left[  \varepsilon
,\Delta\right]  =\frac{1}{n!}\
\end{equation}
and redefining $\epsilon\ln\left[  \mu^{2}/m^{2}\right]  \equiv\epsilon$ \ one
has
\[
J_{n}^{\text{LL}}=\frac{\epsilon^{n+1}}{(n+1)!}.
\]
Then the expansion (\ref{qbar}) \ reads:%
\[
q^{\rm sing}(x)=-\frac{2}{N}\sum_{n=0}^{\infty}~\delta^{(n)}(x)~\left\langle
x^{n}\right\rangle \frac{\epsilon^{n+1}}{(n+1)!}.
\]

The evaluation of the second, two-loop graph with the two propagators
(\ref{prop}) in Fig.\ref{fig1} \ is more complicated due to additional divergency of
the box subgraph. But technically it can be done in the same way therefore we
shall not repeat this discussion.

\end{document}